# Performance Analysis of Decoupled Cell Association in Multi-Tier Hybrid Networks using Real Blockage Environments

Osama Waqar Bhatti*, Haris Suhail*, Uzair Akbar*, Syed Ali Hassan*,
Haris Pervaiz†, Leila Musavian‡ and Qiang Ni†
*School of Electrical Engineering & Computer Science (SEECS), National University of Sciences & Technology (NUST), Islamabad, Pakistan {13beeobhatti, 13beehsuhail, 13beeuakbar, ali.hassan}@seecs.nust.edu.pk
†School of Computing & Communications, Lancaster University, UK.
{h.pervaiz, q.ni}@lancaster.ac.uk
‡School of Computer Science and Electronic Engineering, University of Essex, UK.
leila.musavian@essex.ac.uk

***Abstract*—Millimeter wave (mmWave) links have the potential to offer high data rates and capacity needed in fifth generation (5G) networks, however they have very high penetration and path loss. A solution to this problem is to bring the base station closer to the end-user through heterogeneous networks (HetNets). HetNets could be designed to allow users to connect to different base stations (BSs) in the uplink and downlink. This phenomenon is known as downlink-uplink decoupling (DUDe). This paper explores the effect of DUDe in a three tier HetNet deployed in two different real-world environments. Our simulation results show that DUDe can provide improvements with regard to increasing the system coverage and data rates while the extent of improvement depends on the different environments that the system is deployed in.**

## I. Introduction

A global increase in network traffic demands a shift from the conventional single-tier homogenous networks to multi-tier heterogeneous networks (HetNets). Among the existing techniques, two key enablers of 5G technology are network densification and utilization of higher frequency bands, such as the millimetre wave (mmWave) spectrum [1]-[3]. In the past, mmWave technology was not considered to be feasible for wireless communication due to its higher penetration loss. However, it has been observed that this challenge can be overcome by using highly directional antennas and beamforming [4]-[5]. Moreover, mmWave networks have been shown to be noise-limited rather than interference-limited due to their directional and blockage-sensitive nature [6]. This makes mmWave networks an attractive proposition, at least in integration with traditional ultra high frequency (UHF) networks in 5G HetNets.

User association, a phenomenon where users connect to existing infrastructure, is generally based on the downlink (DL) received signal power [7]. For homogeneous networks, where all the BSs have similar transmit powers, this approach seems feasible. However, a HetNet usually operates on different transmit powers for different tiers, making the aforementioned approach highly inefficient specially for uplink (UL) association. Therefore, DUDe has recently shown to significantly improve the network capacity (especially in the UL) by considering different association criteria for the UL and DL [8].

A decoupled user is defined as a user that has different base stations associated in the forward and reverse channels. Decoupling allows user equipments (UEs) to connect to the best BS in the downlink as well as in the uplink. As a result, decoupling improves the coverage probability of the UE, where a UE is considered to be in coverage if it meets the minimum rate requirement in both the uplink and the downlink.

In the downlink, a UE can connect to any BS depending on proximity of that BS and the UE, the transmit power of the BS, and the offloading bias of that BS. While coupled users are bound to connect to the same BS in the uplink and downlink this may not be ideal as a UE/BS pair that provides the best downlink connection may not necessarily provide the best uplink connection. Decoupling therefore improves the coverage of system essentially by increasing the uplink coverage probability.

### A. Related Work

The authors in [1] and [8] discuss DUDe as an interesting component for HetNets. [8] shows a significant improvement in rate and signal-to-interference plus noise ratio (SINR) when decoupling is applied during user association. The authors in [5]-[7] also show similar results from a theoretical perspective. These however do not discuss the gains of DUDe for mmWave networks in particular.

Effects of DUDe in a two-tier UHF and mmWave deployment was recently studied in [13]. A more complete analytical study on downlink-uplink decoupling for the mmWave-UHF hybrid network is done in [14]. However, this study did not consider the effect of decoupling in different blockage environments and used a simpler two-tier mmWave-UHF system.

Since, for mmWave networks, the path loss depends on links being line-of-sight (LoS), modeling of blockages be-comes important. Analytical modeling of blockages for urban areas via curve fitting techniques have received a lot of attention recently [6]-[9]. However, these models lack the flexibility to be applied in other scenarios of user deployment, e.g. in rural setting. In [10], a line-of-sight (LoS) ball approximation was derived to model the blockages. This blocking model was modified in [11] by adding a LoS probability within the LoS ball in order to more realistically reflect several blockage scenarios.

### B. Contribution and Organization

The main contribution of this paper is to analyze performance of a three-tier hybrid mmWave-UHF network and investigate the gains of DUDe technique in different blockage environments. We use a realistic scenario of a cellular network for different classes of real-world environments, i.e., the National University of Sciences and Technology (NUST) Campus (NC), a sub-urban setting, and the downtown of Chicago city (CC), a denser setting. To the best of our knowledge, this is the first work that assesses the benefits of DUDe in a three-tire mmWave-UHF network deployment for real environments.

The rest of the paper is organized as follows: the system model is presented in Section II. The performance analysis and simulation results are presented in Section III. Finally, the conclusion and final remarks are given in Section IV.

## II. System Model

We consider a three-tier heterogeneous network where the UHF-based macrocells (Mcells), mmWave small cells (Scells) and UHF-based Scells are uniformly distributed in $\mathbb{R}^2$ according to independent homogeneous Poisson Point Processes (PPPs), $\Phi_m$, $\Phi_{s_1}$ and $\Phi_{s_2}$ with densities $\lambda_m$, $\lambda_{s_1}$ and $\lambda_{s_2}$, respectively. Furthermore, $\lambda_{s_2} = \gamma\lambda_s$ and $\lambda_{s_1} = (1-\gamma)\lambda_s$ where $\lambda_s$ is the total intensity of small cells and $0 \leq \gamma \leq 1$ is a parameter used to control the relative intensities of mmWave Scells and UHF Scells. Specifically, a deployment of UHF Mcells overlaid by mmWave and UHF Scells is considered. The user equipments (UEs) are also assumed to be uniformly distributed according to a homogeneous PPP $\Phi_u$ with density $\lambda_u$. The analysis is done for all UEs in a round-robin manner where the BS serving the UE is referred to as the tagged BS.

### A. Path Loss Model

The path loss $L_{mm}(r)$ for the mmWave link, in dB, is modeled as

$$L_{mm}(r) = \begin{cases} \rho + 10\alpha_L \log(r) + \chi_L & \text{if LoS,} \\ \rho + 10\alpha_N \log(r) + \chi_N & \text{otherwise.} \end{cases} \quad (1)$$

In the above equation, $r$ is the radial distance between the transmitter and the receiver while $\chi_L$ and $\chi_N$ are the zero mean log normal random variables (RVs) for LoS and non-line-of-sight (NLoS) mmWave links, respectively, which model the effects of shadow fading. The fixed path loss in $L_{mm}$ is given by $\rho = 20\log\left(\frac{4\pi f_{mm}}{c}\right)$ where $f_{mm}$ is the carrier frequency for mmWave. The path loss exponents in LoS and NLoS mmWave links are denoted by $\alpha_L$ and $\alpha_N$, respectively.

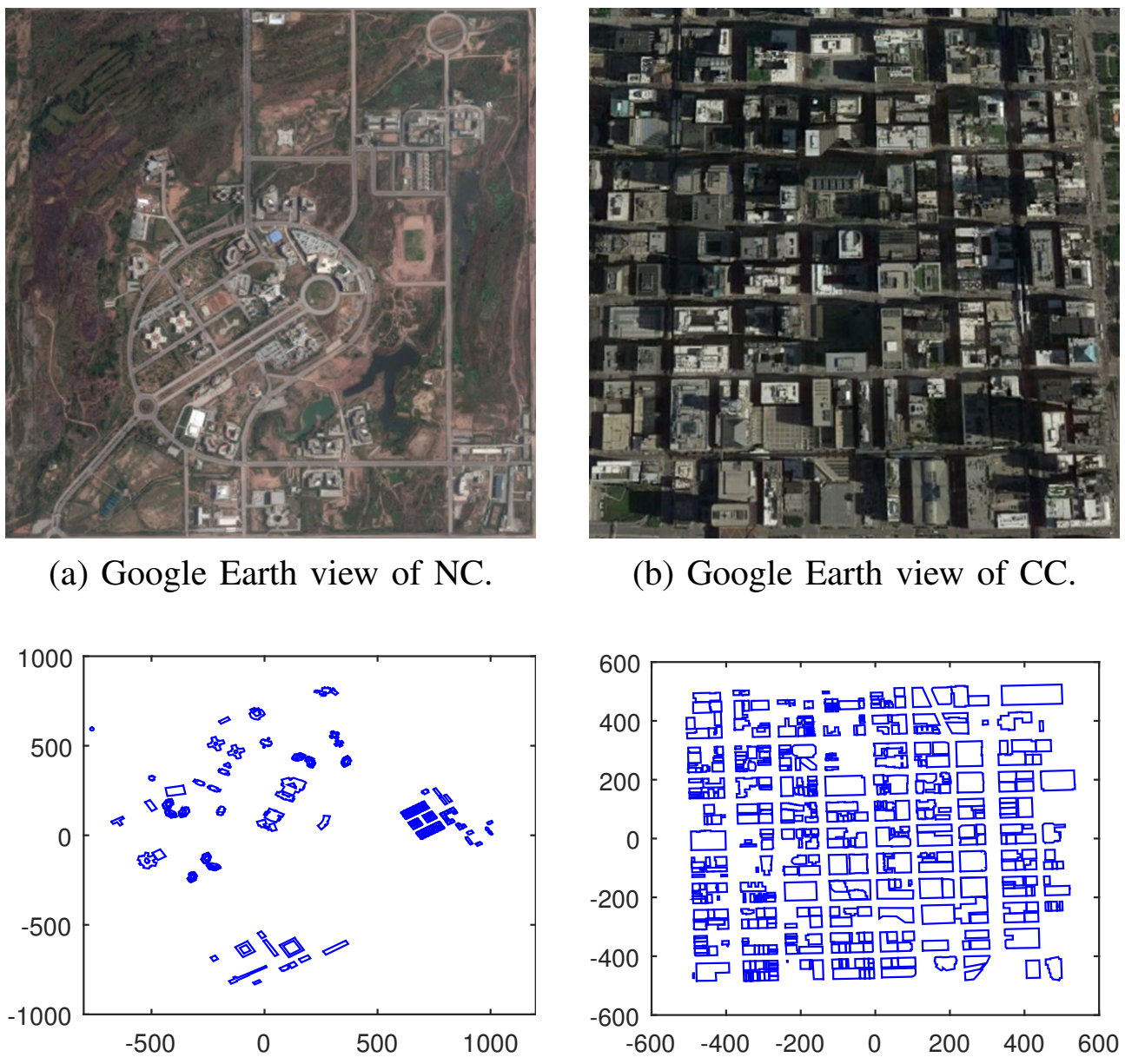

(a) Google Earth view of NC. (b) Google Earth view of CC.

(c) NC extracted building locations. (d) CC extracted building locations.

Fig. 1: Environments under consideration.

Similarly, the path loss for the UHF link, $L_{UHF}(r)$ is given by

$$L_{UHF}(r) = 20\log\left(\frac{4\pi f_{UHF}}{c}\right) + 10\alpha\log(r) + \chi_{UHF}, \quad (2)$$

where $\alpha$ represents the path loss exponent and $\chi_{UHF}$ represents shadow fading in the UHF link.

### B. Blockage Model

We assume a simple stochastic blockage model as proposed in [12] and [15]. Blockages are modeled using a Boolean model of rectangles based on the random shape theory. A user is considered to be in LoS with the following probability

$$p(r) = e^{-\beta r}. \quad (3)$$

Here, $r$ is the distance between the user and the tagged BS and $\beta$ is computed using statistics of the buildings such as density and average size of the blockages in the considered region. It is computed as

$$\beta = \frac{-\rho \ln(1-\kappa)}{\pi A}, \quad (4)$$

where $A$ is the average area of the buildings in the considered region, $\kappa$ is the fraction of area under buildings and $\rho$ is the average parameter of the buildings in the considered region. These parameters are extracted for different real-world environments using the Quantum Geographic Information System (QGIS) software. We use the actual building locations of NC and CC to serve as the test environments. The environments under consideration are depicted in Fig. 1.

### C. User Association

In this paper, both the uplink and downlink user associations are done according to maximum biased received power criterion. The users lying within the considered area are associated with the BS offering the highest (biased) received signal power. Since uplink transmit power of the UEs and the downlink transmit power of the BSs of various tiers is different, this approach will produce a reasonable degree of decoupling.

We assume open access, which allows users to connect to any tier. It is also assumed that each UE is capable of both mmWave and UHF transmission and reception.

Consider the downlink user association scheme first. For a typical UE $i \in \Phi_u$ having a downlink connection with a BS $j \in \Phi_{T_j}$, where $T_j$ is the tier type of the $j^{th}$ BS so that $T_j \in \{m, s_1, s_2\}$, the associated BS $j$ is given as

$$j = arg\max_{k \in \{m,s_1,s_2\}} \frac{P_{t,\text{DL},k} G_{T_k} \beta_{\text{DL},T_k}}{L(d_{ik}, f_{T_k})}. \tag{5}$$

Here, $\beta_{\text{DL},T_k}$ is the downlink bias factor of the tier type of BS $k$, $P_{t,\text{DL},k}$ is the downlink transmit power of the $k^{th}$ BS, $G_{T_j}$ is the maximum antenna gain,$G_{\max}$ of BS $k$, $d_{ik}$ is the distance between the BS, $k$, and the user, $i$ and $f_{T_k}$ is the carrier frequency being used by the BS $k$.

For uplink user association, we follow a similar strategy. For a typical UE, $i \in \Phi_u$ considering a uplink connection with a BS, $j \in \Phi_{T_j}$, where $T_j \in \{m, s_1, s_2\}$, the associated BS $j$ is given as

$$j = arg\max_{k \in \{m,s_1,s_2\}} \frac{P_{t,\text{UL},k} G_{T_k} \beta_{\text{UL},T_k}}{L(d_{ik}, f_{T_k})}, \tag{6}$$

where $P_{t,\text{UL},i}$ is the uplink transmit power of the $i^{th}$ UE and $\beta_{\text{UL},T_k}$ is the uplink bias factor for all users connected to the $k^{th}$ BS. Since mmWave networks are noise-limited, even for higher densities [14], the interference in mmWave Scells can be neglected.

Since the UHF small cells and the macro cells are taken to be sharing the same frequency band, there will be interference between such small cells and macro cells. However, there will be no interference between uplink and downlink transmissions, as uplink and downlink frequency bands are separated. Furthermore, mmWave small cells will produce no interference for the macro cells and UHF small cells.

Consider a user, $i$, associated with a BS $j$ such that $j \in \Phi_l$ where $l \in \{m, s_2\}$. For such a user, downlink SINR is given by the following equation:

$$\text{SINR}_{\text{DL},i} = \frac{P_{r,\text{DL},ij}}{I_{\text{DL},i} + N_{0,i}}, \tag{7}$$

where $I_{\text{DL},i}$ is the interference at the $i^{th}$ user, $N_{0,i}$ is the noise power at the $i_{th}$ user and $P_{r,\text{DL},ij}$ is the power that is received at user $i$ from BS $j$ in the downlink.

The received power is given by

$$P_{r,\text{DL},ij} = \frac{P_{t,\text{DL},j} h_{ij} G_{ij}}{L(d_{ij}, f_{T_j})}. \tag{8}$$

Interference is given by

$$I_{\text{DL},i} = \sum_{q \in \Phi_l \setminus j} \frac{P_{r,\text{DL},iq} h_{iq} G_{iq}}{L(d_{iq}, f_{Tq})}, \tag{9}$$

where $h_{ij}$ is the small Scell fading power gain where we consider Ricean fading with mean $\mu$ and standard deviation $\sigma$ and $G_{ij}$ is the antenna gain between user $i$ and BS $j$. Note that $G_{ij}$ can be ignored in this case as we consider an antenna gain of unity for non-mmWave BSs.

For a user connected to a mmWave small cell in the downlink, the SINR can be approximated by the SNR with reasonable accuracy because of the noise limited nature of the mmWave networks. Thus, $I_{DL,i} = 0$ in eqn (5).

Uplink SINR calculation follows a similar pattern. For a user $i$, associated with a BS $j$ such that $j \in \Phi_l$ where $l \in \{m, s_2\}$, the uplink SINR is given by

$$\text{SINR}_{\text{UL},i} = \frac{P_{r,\text{UL},ij}}{I_{\text{UL},i} + N_{0,i}}, \tag{10}$$

where $P_{r,\text{UL},ij}$ is the uplink received power from user $i$ to BS $j$, $I_{\text{UL},i}$ is the uplink interference that a signal from user $i$ faces at its associated BS. It is given by the following equation

$$I_{\text{UL},i} = \sum_{y \in \Phi_{Iu}} \frac{P_{r,\text{UL},yj} h_{yj} G_{yj}}{L(d_{yj}, f_{Tj})}. \tag{11}$$

$\Phi_{Iu}$ is a set of all users who share the same uplink resource as user $i$. In this paper, we have considered interference from all users connected to a UHF small cell or to a macro cell. As was the case in the forward channel, a user associated to a mmWave Scell in the reverse channel, the interference term in eqn (10) diminishes to zero.

TABLE I: Simulation Parameters

| Symbol | Parameter | Value |
|---|---|---|
| $\lambda_m$ | Mcell BS intensity | 9.5492e-7 |
| $\lambda_s$ | Scell BS intensity | - |
| $\lambda_{s_1}$ | mmWave Scell BS intensity | $(1-\gamma)\lambda_s$ |
| $\lambda_{s_2}$ | UHF Scell BS intensity | $\gamma\lambda_s$ |
| $\gamma$ | Relative Scell intensity control parameter | $0 \le \gamma \le 1$ |
| $\alpha_L$ | mmWave LoS path loss exponent | 2 |
| $\alpha_N$ | mmWave NLoS path loss exponent | 3.3 |
| $\alpha$ | UHF Path loss exponent | 2 |
| $\chi_L$ | LoS mmWave log normal shadowing | $\mu$ = 0, $\sigma$ = 5.2 dB |
| $\chi_N$ | NLoS mmWave log normal shadowing | $\mu$ = 0, $\sigma$ = 7.38 dB |
| $\chi_{UHF}$ | UHF log normal shadowing | $\mu$ = 0, $\sigma$ = 5 dB |
| $f_{mm}$ | Frequency of mmWaves | 73 GHz |
| $f_{UHF}$ | Frequency of UHF | 2.4 GHz |
| $R_{min}$ | Minimum rate required for coverage | 1Mbps |
| $\beta$ | Blocking parameter | 0.0224(CC)<br>0.0014(NC) |
| $\rho_{k,tx}$ | $k^{th}$ tier transmit power | $\rho_{1,tx}$ = 46 dBm<br>$\rho_{2,tx}$ = 30 dBm<br>$\rho_{3,tx}$ = 30 dBm |

## III. Performance Analysis

In this section, we discuss the performance analysis of the system under consideration using simulation results.

The fraction of decoupled users depends on the blocking probability of the environment being considered (Fig. 1). This fraction changes as we vary the density of small cells in the HetNet.

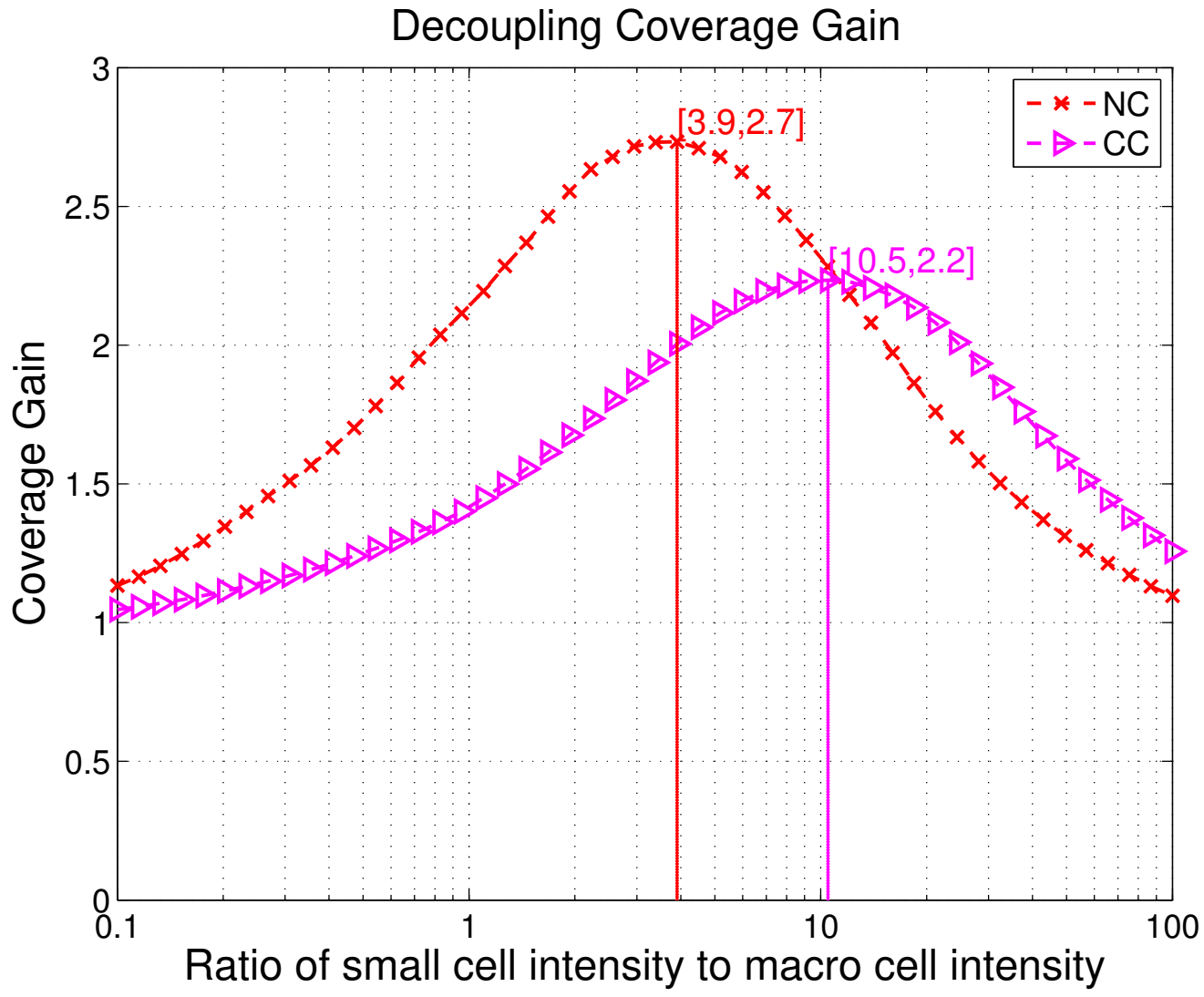

Fig. 2: Coverage gain for different small cell densities

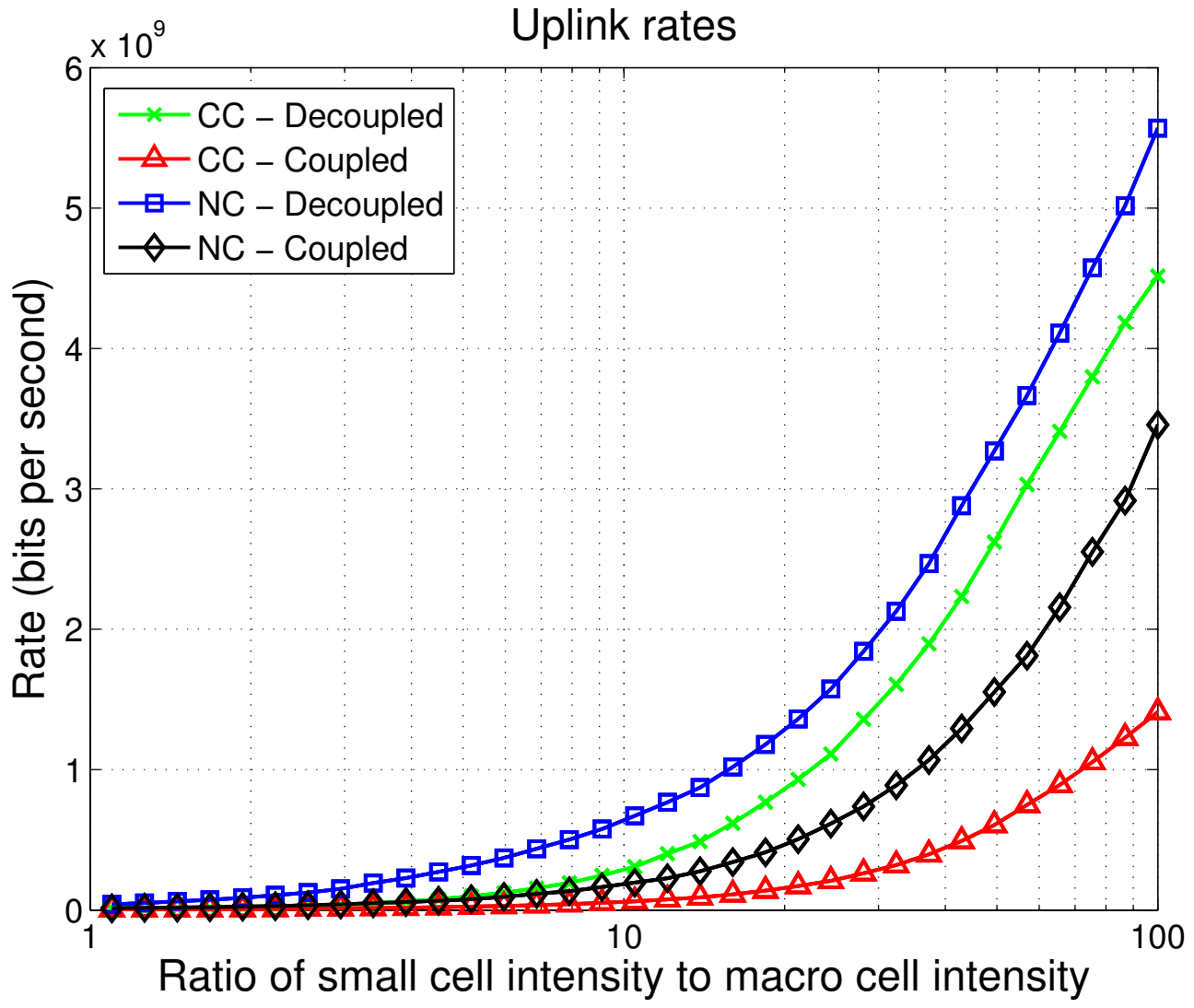

Fig. 3: Uplink data rates with and without decoupling in CC and NC.

We analyze the coverage probability of the system as the ratio of Scells to Mcells changes where $\gamma = 0.3$. The probability of coverage with decoupled downlink and uplink access is referred to as $C_1$, whereas $C_2$ characterizes the overall coverage without DUDe. To analyze the advantage we get by DUDe, we take the ratio $C_1$ to $C_2$, termed as coverage gain. Fig. 2 represents this gain in both CC and NC.

In high blocking environments, such as CC, the coverage gain of decoupling is not as high as in a low blocking environment(NC). This is because at high blocking probability, the chance of establishing a LoS link with the BS decreases. Consequently, the decoupling role of mmWave cells is prohibited as the mmWave path loss becomes unacceptably high. UEs will therefore be left with essentially 2 tiers to choose from, i.e., UHF macro cells and UHF small cells. With fewer uplink BS options to choose from, a UE is less likely to develop a connection that is as good as it would develop in a low blocking environment. Decoupling gain is therefore larger in low blocking environments, as evident in Fig. 2.

Additionally, in NC, the peak decoupling gain occurs at a smaller small cell density than it does in CC. This is due to the smaller role of mmWave BSs in CC. Due to high mmWave path loss and more NLoS links in CC, the UEs will tend to connect to BSs operating on UHF. This effect is the same as reducing the mmWave cell density. The peak coverage, thus, occurs at a higher Scell density in CC. Fig. 3 shows the achievable rates in both the environments with and without decoupling. From Fig. 3, it is evident that DUDe improves system rate in both types of environments. This improvement is due to the ability of decoupled users to connect to BSs that provide the best SINR in the uplink as well as in downlink. Moreover, DUDe allows users to connect to mmWave BSs in the uplink even if a UHF connection is preferred in the downlink. The data rates in CC are markedly lower than those in NC with and without decoupling. This is due to the high blocking probability in CC which makes it less likely for UEs to connect to mmWave Scells.

Because of the higher transmit power of macro cells, the lower path loss of UHF links and the high path loss of mmWave links in CC, UEs in CC are more likely to connect to UHF macro and UHF small cells in the downlink. With DUDe disabled, the UEs will have to connect to the same BSs in the uplink as well. However, with DUDe allowed, the UEs can now connect more to mmWave cells in the uplink, depending on the proximity of the user and the BS. As the Scell BS density increases, the probability of LoS connections rises. Therefore, a decoupled UE can connect to mmWave links in the uplink, significantly improving the data rate.

This effect is not as prominent in NC. Due to the higher probability of LoS connections, UEs are more likely to connect to mmWave cells in the downlink, as well as uplink. So even without decoupling, the number of high speed mmWave links will be high.

Fig. 4 shows how the downlink coverage for the three tiers changes as the Scell BS deployment gets denser. As expected, as the number of Scell BSs increases, the Mcell coverage falls while the Scell coverage rises. Since Scells are larger in number, the UEs tend to connect to Scells rather than Mcells. The downlink bias of mmWave Scells assists in this offloading. The fall in Mcell coverage is also accelerated by the increasing number of UHF Scells, which increases interference in the system. The coverage of mmWave Scell is very low in CC due to greater blocking.

In both environments, the overall coverage falls before rising again. The initial fall is because the decrease in Mcell coverage is greater than the increase in Scell coverage. Beyond a certain Scell density, however, the rise in Scell coverage becomes steeper than the decline in Mcell coverage and the overall probability of coverage escalates.

Fig. 5 shows the uplink coverage of the three tiers at different Scell densities. In both environments, the coverage of Mcells falls, but that of Scells rises with increasing Scell density. In CC, however, the UHF Scells take up the bulk of the UE load whereas in NC, the UEs tend to connect to mmWave cells more often.

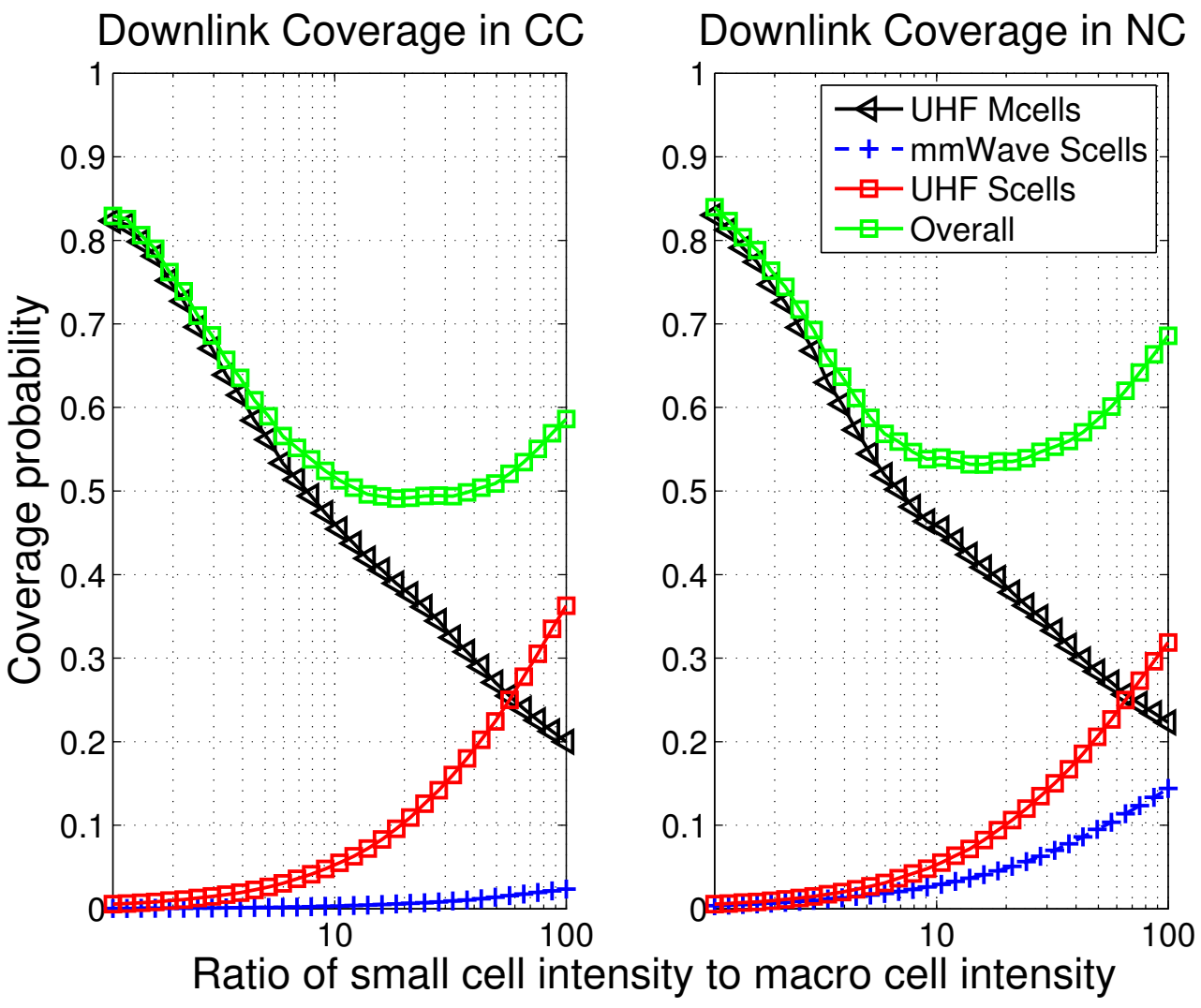

Fig. 4: Downlink coverage in NC and CC.

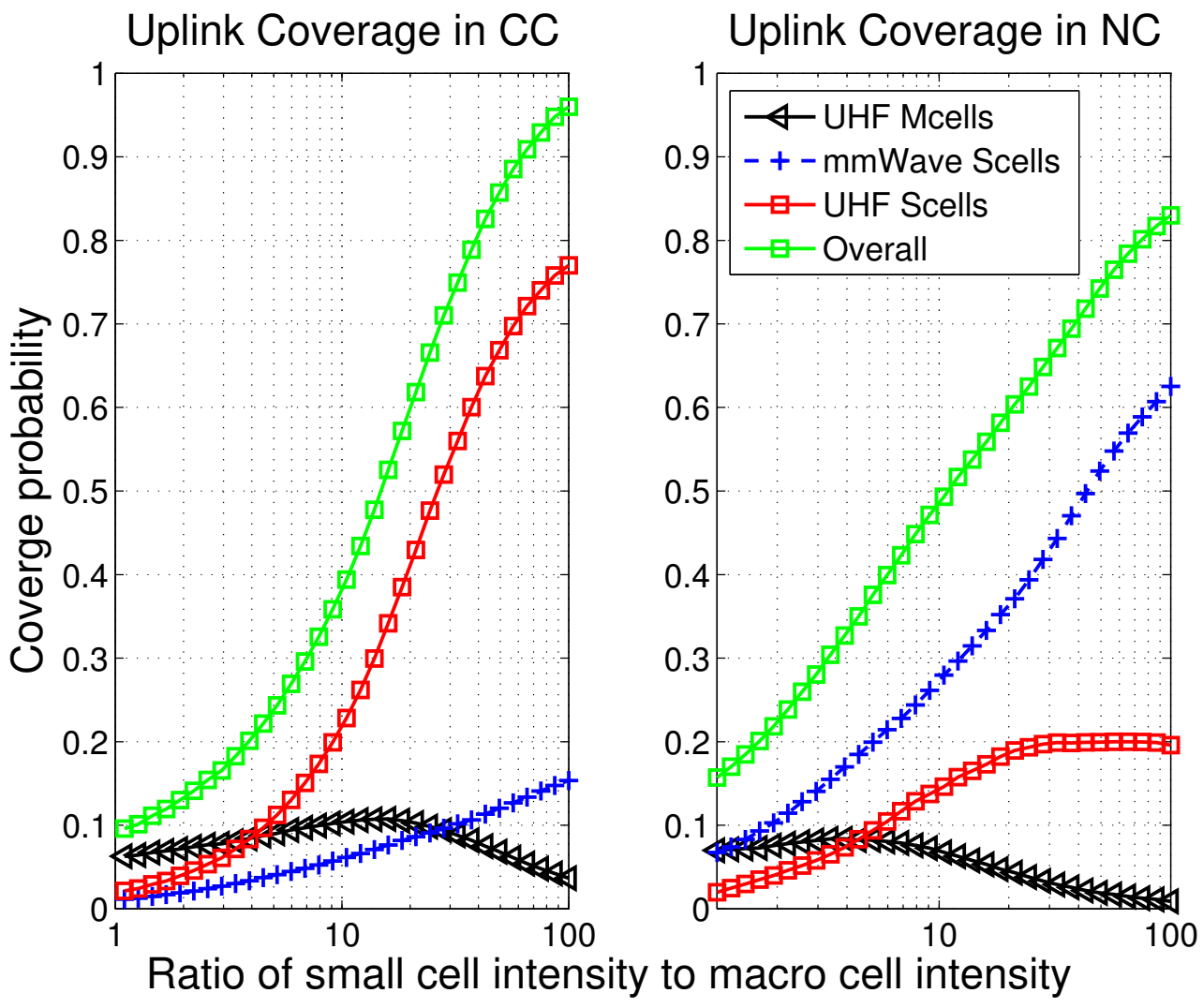

Fig. 5: Uplink coverage in NC and CC.

In Fig. 6, another important perspective is presented. The ratio of mmWave Scells to UHF Scells is varied from the system having all mmWave Scells up to the HetNet consisting of only UHF based Scells. The downlink and uplink data rates are examined both with and without DUDe. Uplink rates are always less than downlink rates because of lower transmit power of the UE. Downlink rates at NC are far greater than that of CC. This is due to the higher mmWave association in NC which allows for greater bandwidth since there are very less LoS links in CC due to greater density of buildings and more multi-paths. The difference in the downlink rates decreases as we go to a HetNet where the Mcells and the Scells operate on the UHF band. This indicates a decrease in the system bandwidth available in the forward path.

We compare the uplink rates in both the environments with and without DUDe. When the system consists only of UHF based Mcells with all the Scells operating on mmWave, we see the greatest difference at Chicago, in the UL data rates. When we allow a user to connect to different base stations in the uplink and downlink, it increases their rates, so as at a certain ratio of the mmWave Scells to UHF based Scells. In short, to get the maximum advantage of decoupling with regard to data rates, CC needs to have a much greater amount of mmWave Scells than UHF based Scells.

However, the case for NC is different. Uplink data rates with and without DUDe are same when only mmWave Scells are deployed. We get the maximum improvements in rates when the ratio is 0.3. This follows from the fact, that there is already a considerable amount of decoupled users in NC as compared to CC. System data rates saturate to a minimum when there is no mmWave Scell deployment. This, thus, is not a good deployment strategy for any kind of environment.

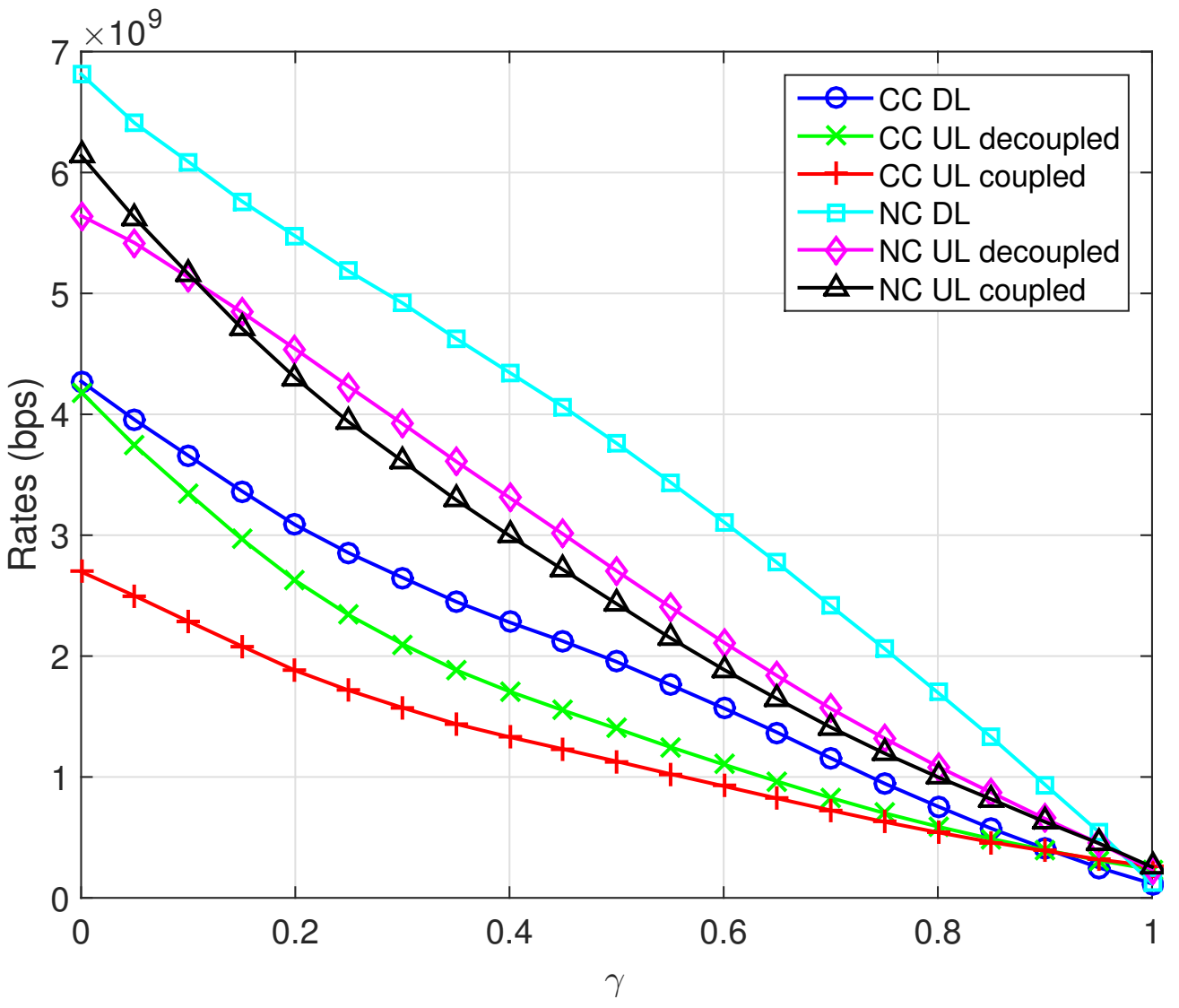

Fig. 6: DL and UL data rates in NC and CC with and without decoupling.

We deduce that mmWave small cells should always be greater than UHF based small cells to get improvements in downlink and uplink rates. In Fig. 7, the fraction of decoupled users is presented for NC and CC. The effect is being analyzed against the ratio of mmWave scells to UHF based Scells. When the system only comprises of mmWave small cells, there are notable amount of decoupled users in the NUST Campus as opposed to the lower number of users who are decoupled in Chicago city which is almost 15%. This is because of the low number of mmWave associations in Chicago due to less number of LoS links. As UHF Scells increase in the system, we see a decline in decoupled users at NC and a gradual increase in CC. This effect takes place till the point where there are equal Scells operating on mmWave and the UHF band. As we deploy more number of Scells which operate on lower frequencies, the coupling increases till there are same fraction of users who are decoupled in both the environments.

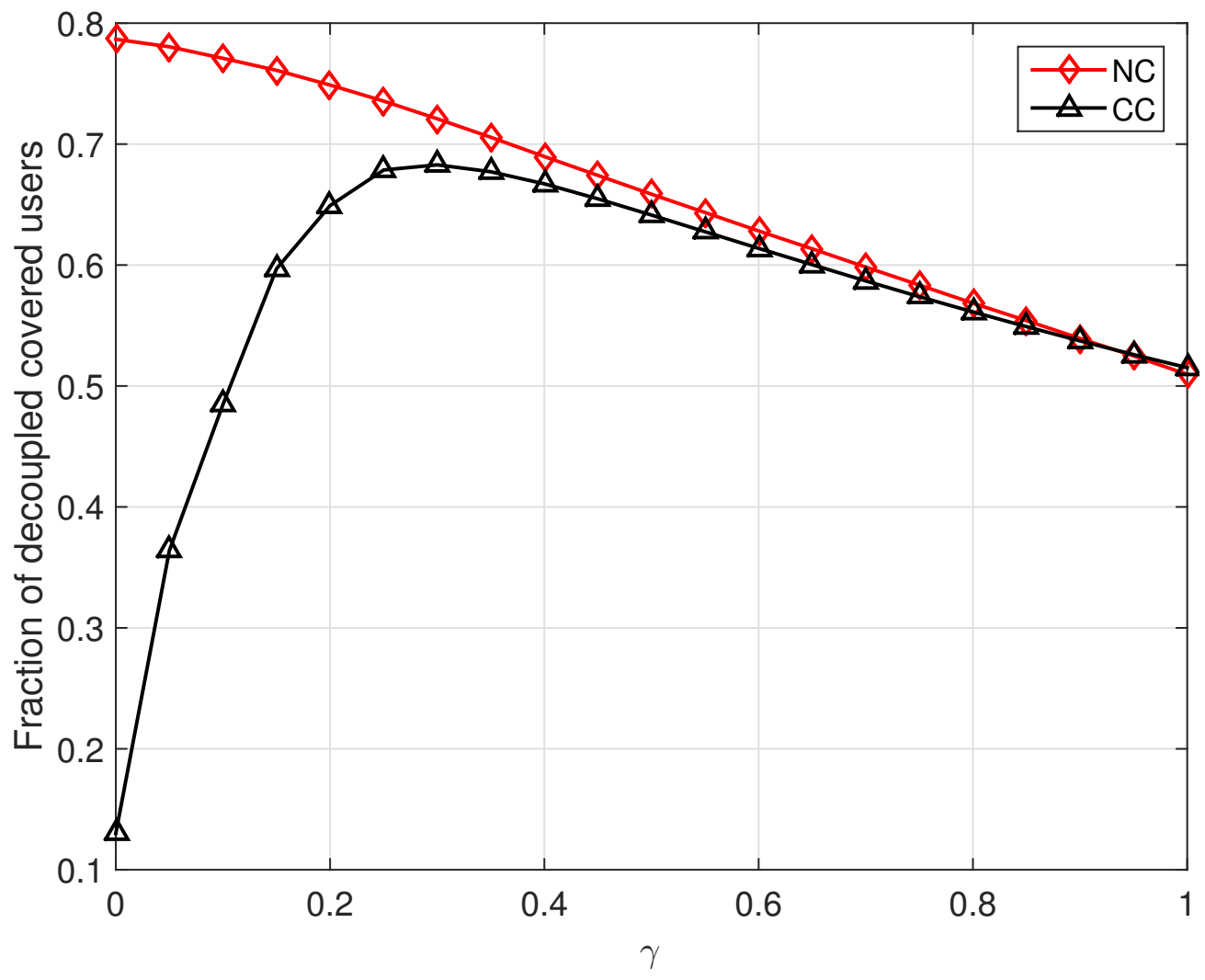

Fig. 7: Fraction of decoupled users in NC and CC.

Integrating with the data rate analysis, we see a trade-off. To get the maximum advantage on rates in an urban environment, we have to have all mmWave Scells(Fig.6) but the fraction of decoupled users would be far less(Fig. 7). Sub-urban environments see maximum decoupling when there are only mmWave small cells but for the maximum improvement on rate, we need to have approximately 30% of small cells operating in the sub-6GHz frequency spectrum.

## IV. Conclusion

In this paper, we have investigated the effect of decoupling uplink and downlink on coverage and data rates in two real environments - a sub-urban environment, such as the NUST Campus, and a dense urban area, i.e., Chicago city. We explored the possibility of various HetNet deployments ranging from one consisting of only UHF band BSs to the one where small cells operate only on mmWave frequencies. It has been observed that there is a larger amount of users which are connected to different BSs in the uplink and downlink in NC than that in CC. However, the decoupling improvement on rate and coverage is greater in CC as compared to NC because of high sensitivity to blockage in the former environment. Simulation results suggest that the DUDe can provide considerable advantage with regard to increasing the system coverage and data rates. User association is currently based on maximum biased received power criterion. Other criteria like minimum rate and minimum spectral efficiency requirement for uplink and downlink separately will be the focus of our future work. Different schemes can be investigated to show which association patterns are optimal for downlink and uplink.